\documentclass[runningheads]{llncs}
\usepackage[T1]{fontenc}
\usepackage{graphicx}
\usepackage[table]{xcolor}
\definecolor{highlight}{RGB}{255, 191, 0}
\definecolor{lightgray}{gray}{0.7} 

\usepackage{subcaption}
\usepackage[mode=buildnew]{standalone} 
\usepackage{url}
\usepackage{xurl} 
\usepackage[]{hyperref}
\usepackage{tabularx}
\usepackage{multirow}
\usepackage{cleveref}
\newcommand{\yesm}{$\bullet$} 
\newcommand{\nom}{} 
\usepackage{xspace}
\newcommand{\nummodels}{17\xspace}
\newcommand{\numdatasets}{4\xspace}

\newcommand{\numlangs}{5\xspace}

\newcommand{\baseLabel}{{b}\xspace}

\newcommand{\standaloneLabel}{{s}\xspace}
\newcommand{\bcbLabel}{BCB\xspace}

\newcommand{\bcbiiLabel}{BCB2\xspace}
\newcommand{\bcbiiiLabel}{BCB3\xspace}
\newcommand{\codenetLabel}{CNet\xspace}
\newcommand{\multipleLabel}{MplE\xspace}
\newcommand{\xcodeevalLabel}{XCE\xspace}
\usepackage{booktabs}
\begin{document}
\title{Candidate Recall in Large-Scale Code-to-Code Retrieval: A Comprehensive Benchmark}
\titlerunning{Candidate Recall in Large-Scale Code-to-Code Retrieval}
%
\author{Leonardo Venuta\inst{1}\orcidID{0009-0007-0328-4001} \and
Francesco Tosoni\inst{1}\orcidID{0000-0001-8457-3866} \and
Paolo Ferragina\inst{1}\orcidID{0000-0003-1353-360X}}
\authorrunning{L. Venuta et al.}
\institute{Sant'Anna School of Advanced Studies, Pisa, Italy\\
\email{\{Leonardo.Venuta, Francesco.Tosoni, Paolo.Ferragina\}@santannapisa.it}
\url{https://www.santannapisa.it/} 
}
%
%
%
\maketitle              
\begin{abstract}
Semantic code search and clone detection are essential for software development, maintenance, and reuse. This paper evaluates the effectiveness, efficiency, and scalability of contemporary deep learning models for first-stage recall in large-scale code-to-code search engines. Benchmarking across multiple programming languages and datasets reveals critical limits in the precision and scalability of these models on Terabyte-scale source-code collections. We present LLM-based code normalisation and query-rewriting schemes that yield significant gains in precision for lower-performing models. Our results question the sustainability of resource-constrained deployment and the assumed robustness of current code-specialised LLMs across datasets. We conclude with actionable insights for building scalable, efficient code-retrieval systems.
\keywords{Semantic code search \and
Code embeddings \and
Two-stage retrieval \and
Benchmark \and
Query rewriting
}
\end{abstract}
\section{Introduction}
As software repositories grow to unprecedented scale (the Software Heritage (SWH) archive~\cite{swh_cacm} held over 2 petabytes across 421 million projects as of January 2026~\cite{swh_report_2025}), much of the code developers write already exists elsewhere or is semantically redundant. Code search, which retrieves relevant snippets from natural-language or code queries, has thus become indispensable to software development and maintenance, underpinning code completion, synthesis, traceability, and vulnerability detection.
A fundamental tension separates the accuracy of semantic deep learning (DL) models, which incur steep inference costs~\cite{two-stage-paradigm-2023}, from the speed of classical IR metrics (BM25~\cite[\S11.4.3]{ir_book}\cite[\S3.2.1]{aipowered-book}, TF-IDF~\cite[\S6.2.2]{ir_book}\cite[\S3.1.6]{aipowered-book}, and Jaccard~\cite[\S19.6]{ir_book}), which deliver millisecond inference but rely on surface-level keyword matching.
Large-scale, end-to-end retrieval systems typically resolve this trade-off through a {\em two-stage recall-then-rerank} paradigm: a lightweight embedding model first narrows the corpus $\mathcal{C}$ to a small candidate set, after which a high-accuracy cross-encoder reranks each $\langle q, c_i\rangle$ pair to surface the snippets most similar to $q$. TOSS~\cite{two-stage-paradigm-2023} formalised this architecture as a baseline for code-to-code search. The pipeline hinges on first-stage recall, its principal bottleneck: any relevant snippet missed here cannot be recovered by even the most sophisticated index or reranker.
\paragraph{Our Contribution.}
Despite its centrality, the recall stage lacks a systematic analysis of its state of the art. We close this gap with the first large-scale empirical study of DL model selection for code-to-code search, evaluating \nummodels{} diverse transformers across \numdatasets{} datasets and \numlangs{} programming languages. The study yields three contributions:
\begin{enumerate}
    \item A large-scale benchmark of \nummodels{} encoder and decoder models, drawn from prior work~\cite{two-stage-paradigm-2023,comparison-pretrained-source-code-2023} and forward snowballing, spanning diverse architectures, parameter scales, and training paradigms. We measure efficacy (Precision@$k$~\cite[\S8.3]{ir_book}, NDCG~\cite[\S8.4]{ir_book}) and efficiency (KB/sec throughput) across \numdatasets{} datasets, \numlangs{} languages, and 300k+ code snippets. The benchmark has spanned 920 experimental runs ($644$ retrieval runs: 23 models $\times$ $14$ dataset $\times$ $2$ distance metrics; plus $276$ rewriting runs: 23 models $\times$ 3 datasets $\times$ 4 variants), thus requiring $\sim$403 GPU hours, reflecting the scale of the evaluation. Undoubtedly, the first study at this scale, and the first to apply the MultiPL-E dataset~\cite{multiple} to code-to-code retrieval.
    
    \item Empirical findings across five languages (Python, Java, JavaScript, C++, and C\#) and four varied datasets that reveal a quality--scalability dichotomy. On retrieval quality alone, Qwen3 Embedding and CodeXEmbed lead across languages and datasets, with specialised embedders surpassing far larger general-purpose LLMs; yet these gains entail prohibitive computational costs at scale. Light\-wei\-ght models recover an order-of-magnitude throughput gain but sacrifice precision steeply (up to 80 points on xCodeEval), underscoring the need for task- and resource-specific model selection. This throughput gap between lightweight encoders (e.g., StarEncoder) and large language models impedes the adoption of LLM-embedding indexes on medium-sized datasets by small- and medium-sized enterprises (SMEs) and academic researchers.
    \item A code-rewriting analysis showing LLM-based style normalisation lifts recall by up to 29\% for weaker, style-sensitive models (e.g., Code Llama), whereas top performers stay robust to variations in style, comments, and identifiers.
\end{enumerate}

\noindent Our results confirm that model selection must be task- and context-specific, requiring a careful balance between throughput, accuracy, and language specialisation. Further, they reinforce the conclusion that a hybrid two-stage pipeline: combining a fast, compact encoder for candidate retrieval with a powerful LLM for reranking, remains the most effective architecture \cite{two-stage-paradigm-2023}. An additional takeaway from our investigation is that, despite the availability of numerous general-purpose and code-specific embedding models in the literature, a universally optimal solution has yet to emerge, even for widely used programming languages such as Python, Java, JavaScript, C++, and C\#. In particular, when practical scalability requirements are taken into account and GPU resources are limited (as in most academic scenarios), achieving higher throughput often entails a substantial loss of retrieval precision, which in turn leads to serious limitations in the design of effective code retrieval systems. These findings (with implementation released at \href{https://github.com/leeeov4/code2code_benchmark}{github.com/leeeov4/code2code\_benchmark}) highlight the need for further research and large-scale empirical validation to better understand the trade-offs between efficiency, accuracy, and language-specific performance in those settings. For a more comprehensive discussion of actionable guidelines and future research directions, we refer readers to \Cref{sec:future}.

\begin{figure}
  \centering
  \includegraphics[width=.99\linewidth]{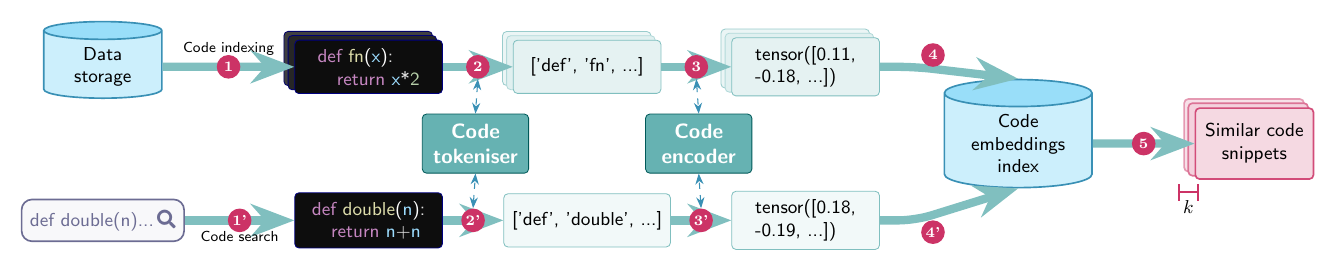}
  \caption{End-to-end top-$k$ code retrieval architecture. Offline stages (1--4) build a DL-based embedding index; online ones (1'--4') search it to retrieve the $k$ snippets most similar to query $q$. To isolate model performance from lossy-compression effects, we compute ground-truth nearest neighbours via sequential scans.}\label{fig:flow-horizontal}
\end{figure}
\section{Background}
\label{sec:background}
Deep learning has markedly advanced semantic code search for both language-to-code and code-to-code retrieval. Generation Augmented Retrieval (GAR)~\cite{rewriting-code}, e.g., rewrites queries before they reach the vector index. Classical IR methods (BM25~\cite[\S11.4.3]{ir_book}\cite[\S3.2.1]{aipowered-book}, TF-IDF~\cite[\S6.2.2]{ir_book}\cite[\S3.1.6]{aipowered-book}, Jaccard~\cite[\S19.6]{ir_book}) deliver millisecond-scale inference but match only at the surface, capping their accuracy. Neural bi-encoders, instead, map queries and snippets independently into a shared vector space~\cite{two-stage-paradigm-2023}, letting precomputed embeddings be indexed for efficient ANN search \ref{fig:flow-horizontal}. Cross-encoders process query--candidate pairs jointly, trading scalability for higher accuracy. This split motivates the two-stage recall-then-rerank paradigm (TOSS~\cite{two-stage-paradigm-2023}): a lightweight bi-encoder (optionally augmented with BM25) first narrows the candidate pool, then a cross-encoder reranks the subset into the final results. Any snippet missed during recall cannot be recovered downstream.
Existing benchmarks paint only a partial picture. TOSS~\cite{two-stage-paradigm-2023} evaluated CodeBERT~\cite{codebert2020} and GraphCodeBERT~\cite{graphcodebert2021} on Python CodeSearchNet~\cite{codesearchnet}; a broader study~\cite{comparison-pretrained-source-code-2023} benchmarked 19 models across 13 tasks, finding UniXcoder~\cite{UniXcoder2022} and GraphCodeBERT best on POJ-104, a C-only dataset. Both omit newer architectures (CodeSage-V2~\cite{codesagev2}, CodeXEmbed~\cite{codexembed}, Nomic Embed~\cite{nomic_embed}, Code Llama~\cite{codellama2024}, Qwen~\cite{qwen3}), cover at most two datasets per language, and disregard throughput. Embedding models are sensitive to superficial style variations too~\cite{rewriting-code,sacl-textual-bias}, which bears on plagiarism detection and vulnerability analysis.
\section{Problem Statement and Research Questions}
\label{sec:problem}
Given a codebase $\mathcal{C}$ and a query snippet $q$, the {\em code search problem} retrieves the top-$k$ snippets from $\mathcal{C}$ most semantically similar to $q$. Following TOSS~\cite{two-stage-paradigm-2023}, we focus on the first-stage recall function $R(q,\mathcal{C})$, which must be effective (high recall across diverse queries and languages) and efficient (scalable to massive code archives). Our primary research questions are as follows:
\begin{itemize}
    \item[RQ$_1$:] How effective are contemporary code embedding models at first-stage recall, and which architectural factors (encoder type, size, pre-training data) most shape performance?
    \item[RQ$_2$:] What throughput and infrastructure bottlenecks dominate when deploying these models over very large code-snippet collections?
    \item[RQ$_3$:] To what extent do semantics-preserving transformations (variable renaming, comment insertion and deletion) affect retrieval efficacy, and which rewriting strategies are most promising?
\end{itemize}
\noindent We address RQ$_1$ in \Cref{sec:model,sec:dataset,sec:isolation}, evaluating \nummodels{} models across \numdatasets{} datasets and \numlangs{} programming languages using Precision@$k$~\cite[\S8.3]{ir_book} and NDCG~\cite[\S8.4]{ir_book}. For RQ$_2$, we show in \Cref{sec:isolation} that embedding rate is scale-critical (e.g., BigCloneBench's 6M methods~\cite{bigclone}), making IR methods or lossy embedding compression~\cite{gaurav-survey2023,binary-ternary-survey2026}\cite[\S13.7]{aipowered-book} more practical for large corpora. For RQ$_3$ in \Cref{sec:rewriting}, we examine LLM- and rule-based rewriting applied to queries and/or codebase.
\begin{table*}[t]
\centering
\setlength{\tabcolsep}{3pt}
\begin{tabular}{p{16mm}p{20mm}r|ccc|ccc|rrr}
\toprule
\textbf{Group} & \multicolumn{2}{c|}{\textbf{Model}} & \textbf{Y} & \textbf{T} & \textbf{A} & \textbf{J/P} & \textbf{JS} & \textbf{C} & \textbf{\#P} & \textbf{Dim.} & KB/s\\
\midrule
\multirow{4}{17mm}{Early Bimodal Encoders}
    & CodeBERT      &\cite{codebert2020} & '20 & G & E & \yesm & \yesm & \yesm & 125M  & 768 & 105 \\
\noalign{\vspace{2pt}}\arrayrulecolor{lightgray}\cline{2-12}\arrayrulecolor{black}\noalign{\vspace{2pt}}
    & G-CodeBERT &\cite{graphcodebert2021} & '21 & G & E & \yesm & \yesm & \yesm & 125M & 768 & 125 \\
\noalign{\vspace{2pt}}\arrayrulecolor{lightgray}\cline{2-12}\arrayrulecolor{black}\noalign{\vspace{2pt}}
    & CoTexT        &\cite{cotext} & '21 & G & E/D & \yesm & \yesm & \nom & 220M & 768 & 116 \\
\noalign{\vspace{2pt}}\arrayrulecolor{lightgray}\cline{2-12}\arrayrulecolor{black}\noalign{\vspace{2pt}}
    & SPT-Code      &\cite{sptcode} & '21 & G & E & \yesm & \yesm & \nom & 262M & 768 & 165 \\

\midrule
\multirow{6}{17mm}{Unified Encoder-Decoders}
    & PLBART        &\cite{plbart} & '21 & G & E/D & \yesm & \nom & \nom  & 140M & 768 & 98 \\
\noalign{\vspace{2pt}}\arrayrulecolor{lightgray}\cline{2-12}\arrayrulecolor{black}\noalign{\vspace{2pt}}
    & \multirow{2}{*}{CodeT5} & \multirow{2}{*}{\cite{codet52021}} & \multirow{2}{*}{'21} & \multirow{2}{*}{G} & \multirow{2}{*}{E/D} & \multirow{2}{*}{\yesm} & \multirow{2}{*}{\yesm} & \multirow{2}{*}{\yesm} & 220M & \multirow{2}{*}{768} & 69 \\
    & & & & & & & & & 770M & & 21 \\
\noalign{\vspace{2pt}}\arrayrulecolor{lightgray}\cline{2-12}\arrayrulecolor{black}\noalign{\vspace{2pt}}
    & UniXcoder     &\cite{UniXcoder2022} & '22 & G & E/D & \yesm & \yesm & \yesm & 125M & 768 & 107 \\
\noalign{\vspace{2pt}}\arrayrulecolor{lightgray}\cline{2-12}\arrayrulecolor{black}\noalign{\vspace{2pt}}
    & \multirow{2}{*}{CodeT5+} & \multirow{2}{*}{\cite{codet5+}} & \multirow{2}{*}{'23} & \multirow{2}{*}{G} & \multirow{2}{*}{E/D} & \multirow{2}{*}{\yesm} & \multirow{2}{*}{\yesm} & \multirow{2}{*}{\yesm} & 110M & 768 & 128 \\
    & & & & & & & & & 220M & 1024 & 129 \\

\midrule
\multirow{7}{17mm}{Code Embedders}
    & StarEncoder   &\cite{the-stack-v2} & '23 & E & E & \yesm & \yesm & \yesm & 125M & 768 & 140 \\
\noalign{\vspace{2pt}}\arrayrulecolor{lightgray}\cline{2-12}\arrayrulecolor{black}\noalign{\vspace{2pt}}
    & \multirow{2}{*}{CodeXEmbed} & \multirow{2}{*}{\cite{codexembed}} & \multirow{2}{*}{'23} & \multirow{2}{*}{E} & \multirow{2}{*}{E} & \multirow{2}{*}{\yesm} & \multirow{2}{*}{\yesm} & \multirow{2}{*}{\yesm} & 400M & \multirow{2}{*}{1024} & 28 \\
    & & & & & & & & & 2B & & 7 \\
\noalign{\vspace{2pt}}\arrayrulecolor{lightgray}\cline{2-12}\arrayrulecolor{black}\noalign{\vspace{2pt}}
    & \multirow{2}{*}{CodeSage-V2} & \multirow{2}{*}{\cite{codesagev2}} & \multirow{2}{*}{'23} & \multirow{2}{*}{E} & \multirow{2}{*}{E} & \multirow{2}{*}{\yesm} & \multirow{2}{*}{\yesm} & \multirow{2}{*}{\yesm} & 356M & \multirow{2}{*}{2048} & 45 \\
    & & & & & & & & & 1.3B & & 15 \\
\noalign{\vspace{2pt}}\arrayrulecolor{lightgray}\cline{2-12}\arrayrulecolor{black}\noalign{\vspace{2pt}}
    & CR-Embed &\cite{CodeRankEmbed} & '24 & E & E & \yesm & \yesm & \nom & 137M & 768 & 67 \\
\noalign{\vspace{2pt}}\arrayrulecolor{lightgray}\cline{2-12}\arrayrulecolor{black}\noalign{\vspace{2pt}}
    & Nomic E.C. &\cite{nomic_embed} & '25 & E & E & \yesm & \yesm & \nom & 7B & 3584 & 2 \\

\midrule
\multirow{6}{17mm}{Code Decoder-Only LLMs}
    & \multirow{2}{*}{Granite Code} & \multirow{2}{*}{\cite{granite_code_models2024}} & \multirow{2}{*}{'24} & \multirow{2}{*}{G} & \multirow{2}{*}{D} & \multirow{2}{*}{\yesm} & \multirow{2}{*}{\yesm} & \multirow{2}{*}{\yesm} & 3B & \multirow{2}{*}{2560} & 9 \\
    & & & & & & & & & 8B & & 1 \\
\noalign{\vspace{2pt}}\arrayrulecolor{lightgray}\cline{2-12}\arrayrulecolor{black}\noalign{\vspace{2pt}}
    & Code Llama        &\cite{codellama2024} & '24 & G & D & \yesm & \yesm & \yesm & 7B & 4096 & 2 \\
\noalign{\vspace{2pt}}\arrayrulecolor{lightgray}\cline{2-12}\arrayrulecolor{black}\noalign{\vspace{2pt}}
    & Qwen3 Coder      &\cite{qwen3} & '25 & G & D & \yesm & \yesm & \yesm & 30B & 2048 & 3 \\
\noalign{\vspace{2pt}}\arrayrulecolor{lightgray}\cline{2-12}\arrayrulecolor{black}\noalign{\vspace{2pt}}
    & \multirow{2}{*}{Qwen3 Emb.} & \multirow{2}{*}{\cite{qwen3_emb}} & \multirow{2}{*}{'25} & \multirow{2}{*}{G} & \multirow{2}{*}{D} & \multirow{2}{*}{\yesm} & \multirow{2}{*}{\yesm} & \multirow{2}{*}{\yesm} & 600M & 1024 & 74 \\
    & & & & & & & & & 8B & 4096 & 9 \\
\bottomrule
\end{tabular}
\caption{Selected models, grouped by architecture, and group-sorted by year (Y).\label{tab:model-overview}}
\end{table*}
\section{Model Selection}
\label{sec:model}
We collated models from prior benchmarks~\cite{two-stage-paradigm-2023,comparison-pretrained-source-code-2023} and used forward snowballing on Google Scholar (keywords: `code embedding', `code retrieval', `code search', `bi-encoder'). This yielded \nummodels{} models (125M to 30B parameters), selected on four criteria: (i)~\textit{architectural paradigm}: bi-encoders (e.g., CodeRankEmbed~\cite{CodeRankEmbed}), encoder-decoders (e.g., CodeT5~\cite{codet52021}), and decoder-only LLMs in embedding mode (e.g., Code Llama~\cite{codellama2024}, Qwen3~\cite{qwen3}); (ii)~\textit{model scale}: favouring the base variant of each family; (iii)~\textit{provenance}: open-source weights, code-specialised training, and varied organisational origins; and (iv)~\textit{prior performance}: from~\cite{comparison-pretrained-source-code-2023} we kept models exceeding 85\% accuracy (PLBART, CoTexT, CodeT5, SPT-Code, UniXcoder, CodeBERT, GraphCodeBERT), discarding DOBF, SynCoBERT, and TreeBERT due to unavailable weights.
We excluded classical IR baselines (BM25, TF-IDF, Jaccard). As~\cite{two-stage-paradigm-2023} establishes, deep learning consistently outperforms rule-based matching for code-to-code retrieval, where semantic similarity requires deeper analysis than surface lexical overlap. Choosing optimal IR metrics to augment neural embeddings is orthogonal to first-stage recall and belongs to later reranking analyses~\cite{two-stage-paradigm-2023}.
\Cref{tab:model-overview} reports each model's release year; target~(T): embedding/retrieval~(E) or generic~(G); architecture~(A): encoder~(E), decoder-only~(D), or encoder-decoder~(E/D); language support: Java and Python~(J/P), JavaScript~(JS), C++ and C\#~(C); parameter count (\#P); embedding dimension; inference throughput. Early encoders (2020--2023) uniformly produce 768-dimensional embeddings (110M--262M parameters). Code-specialised embedders (2023--2024) continue to grow beyond 1024 dimensions, with Nomic Embed Code as an outlier at 7B parameters and 3584 dimensions. Decoder-only models (2024--2025) show the largest and most varied dimensions, marking a paradigm shift in embedding architecture. 
The table groups models into four families: early bimodal encoders (MLM/denoising, often using AST or NL-PL data); unified encoder-decoders; code-specialised embedders (contrastive learning, Matryoshka representations); and general code LLMs adapted for retrieval. By spanning architectures, our benchmark offers comprehensive insight into the efficacy and efficiency of modern code embeddings.
\section{Dataset Selection}
\label{sec:dataset}
\begin{table}[t]
\small
\centering
\newcolumntype{P}{>{\centering\arraybackslash}p{4.5mm}} 

\begin{tabular}{l l P P P P P c c c c r r}
\toprule
\textbf{Dataset} & & \textbf{Ja} & \textbf{Py} & \textbf{Cp} & \textbf{Cs} & \textbf{Js} & \textbf{$k$} & \textbf{Min} & \textbf{Avg} & \textbf{Max} & {\bf \#queries} & {\bf \#targets} \\
\midrule
\bcbLabel &~\cite{bigclone2} & $\bullet$ & & & & & 50 & 50 & 95.7 & 264 
          & 2\,156 & 12\,490 \\
\midrule
\codenetLabel &~\cite{codenet} & $\bullet$ & $\bullet$ & $\bullet$ & & & 50 & 299 & 299 & 299 
          & 750 & 224\,250 \\
\midrule
\multipleLabel &~\cite{multiple} & $\bullet$ & & $\bullet$ & $\bullet$ & $\bullet$ & 20 & 20 & 42.7 & 50 
          & 258 & 17\,885 \\
\midrule
\xcodeevalLabel &~\cite{xCodeEval} & $\bullet$ & $\bullet$ & $\bullet$ & $\bullet$ & $\bullet$ & 50 & 50 & 82.1 & 100 
          & 672 & 58\,667 \\
\bottomrule
\end{tabular}
\caption{For each dataset, we show the included language (Ja: Java, Py: Python, Cp: C++, Cs: C\#, Js: JavaScript), the number of positive snippets per query (Min/Avg/Max), and the total number of queries and target snippets.\label{tab:datasets}}
\end{table}
We evaluate all selected models on four established datasets across \numlangs{} programming languages, chosen for their language coverage, prevalence in recent literature, and availability of ground truth. \Cref{tab:datasets} reports total code snippets in the query and target sets across all considered languages.
To gauge retrieval quality, the cutoff $k$ never exceeds the number of relevant answers per query; queries with fewer than $k$ relevant snippets are discarded. We exclude extremely large-scale codebases like CodeSearchNet~\cite{codesearchnet} due to prohibitive embedding generation costs: \cite{rewriting-code} estimated that embedding its 1,005,474 queries would take over two months. CodeSearchNet also focuses on natural-language code retrieval, which lies outside our scope.
\paragraph{BigCloneBench (BCB) v2.}
We use BCB v2~\cite{bigclone2} as a rigorous, large-scale, manually validated Java benchmark. A code-clone pair is a triple $\left(f_1, f_2, \phi \right)$, where $f_1$ and $f_2$ are method-level snippets and $\phi$ specifies one of four clone types~\cite{code_types_2}:
(i) Type-1 (48,116 pairs): syntactically exact clones differing only in whitespace, layout, and comments;
(ii) Type-2 (4,234 pairs): lexical clones extending Type-1 with variations in identifier names and literal values;
(iii) Type-3 (109,446 pairs): structural clones extending Type-2 with statement-level additions or deletions; and
(iv) Type-4 (8,450,204 pairs): syntactically dissimilar snippets implementing identical functionality.
These labels assess how embedding models handle rising syntactic and structural variance. We exclude Type-4 clones due to documented ground-truth reliability issues, specifically a high false-positive rate~\cite{bigclone-harmful}.
We frame clone detection as a search task. For each clone type, snippets of that type form a separate target corpus. Every distinct function in at least one clone relationship of that type serves as a query. For query $q$, a candidate snippet $c$ is a true positive \emph{iff} the pair $\langle q, c \rangle$ is explicitly labelled as a clone of that type. The target set comprises all functions appearing in at least one clone pair of that type, either as $q$ or $c$. As noted, we discarded queries with fewer than $k$ clones.
\paragraph{CodeNet.}
To evaluate multi-language retrieval using functional correctness as ground truth, we use IBM's CodeNet dataset\footnote{\url{https://github.com/IBM/Project_CodeNet}}. Following the authors' curation (retaining only accepted submissions and removing near-duplicates), all solutions to a problem are functionally equivalent. We derive a retrieval benchmark for C++, Java, and Python, selecting 250 problems per language, each with exactly 300 accepted solutions. We sample uniformly to balance CodeNet's dense solution sets, avoiding skew from disproportionate submission pools while keeping a consistent candidate space across languages. For each problem, one submission is chosen at random as the query; the remaining 299 equivalent solutions form the ground truth, while all solutions from the other 249 problems act as negatives.
\paragraph{MultiPL-E.}
MultiPL-E\footnote{\url{https://huggingface.co/datasets/nuprl/MultiPL-E}} was originally designed to port unit-test-driven code-generation benchmarks to new languages. We leverage its HumanEval translation component to construct a multi-language retrieval benchmark grounded in functional equivalence. MultiPL-E is widely used to evaluate LLM code-generation capabilities~\cite{qwen3,granite_code_models2024,codellama2024}. We evaluate retrieval across Java, C\#, C++, and JavaScript, using candidate solutions generated by StarCoder2~\cite{the-stack-v2}.
The original HumanEval benchmark contains 164 problems, each with at most 50 candidate solutions. We select one accepted solution per problem at random as the query $q$, while the candidate space comprises all translations within the same target language. A retrieved snippet $c$ is a true positive \emph{iff} it stems from the same HumanEval problem as $q$ and passes all associated unit tests, testing a model's ability to discriminate functionally correct implementations from incorrect ones. We use a smaller threshold $k=20$ here (instead of $k=50$) to ensure enough queries contain at least $k$ positive snippets.
\paragraph{xCodeEval.}
Prior work validates xCodeEval for language- and code-to-code retrieval, proving it remains challenging even for models like GPT-3.5 Turbo~\cite{xCodeEval}. Each problem includes both positive (accepted) and negative (incorrect) solutions. We selected snippets in C++, C\#, Java, JavaScript, and Python, discarding the negatives and treating accepted solutions as target sets. We designate the official ``reference'' solutions as queries, so that the number of queries equals the number of unique problems per language. The candidate space for a language comprises all its accepted solutions. A retrieved snippet $c$ is a true positive \emph{iff} it shares the same problem ID as query $q$.
\renewcommand{\arraystretch}{0.6} 
\begin{table}[h]
\setlength{\tabcolsep}{3.5pt}
\centering
\small
\begin{tabular}{
  p{17mm} r |
  *{1}{r!{\color{lightgray}\vrule}}r|          
  *{4}{r!{\color{lightgray}\vrule}}r|         
  *{1}{r!{\color{lightgray}\vrule}}r|          
  *{2}{r!{\color{lightgray}\vrule}}r|          
  *{2}{r}           
}
\toprule

 Language &
 & \multicolumn{2}{c|}{Py} 
 & \multicolumn{5}{c|}{Java} 
 & \multicolumn{2}{c|}{JS} 
 & \multicolumn{3}{c|}{C++} 
 & \multicolumn{2}{c}{C\#}
 \\

Dataset & 

& \rotatebox{90}{\bf \codenetLabel} & \rotatebox{90}{\bf \xcodeevalLabel} 

& \rotatebox{90}{\bf \bcbiiLabel} & \rotatebox{90}{\bf \bcbiiiLabel} & \rotatebox{90}{\bf \codenetLabel} & \rotatebox{90}{\bf \multipleLabel} & \rotatebox{90}{\bf \xcodeevalLabel} 

& \rotatebox{90}{\bf \multipleLabel} & \rotatebox{90}{\bf \xcodeevalLabel} 

& \rotatebox{90}{\bf \multipleLabel} & \rotatebox{90}{\bf \codenetLabel} & \rotatebox{90}{\bf \xcodeevalLabel}

& \rotatebox{90}{\bf \multipleLabel} & \rotatebox{90}{\bf \xcodeevalLabel}

\\ 

\midrule

C.BERT & \baseLabel
& 28 & 15 & {\bf 98} & 55 & 9 & 80 & 5 & 75 & 20 & 75 & 8 & 5 & 71 & 9 \\

\addlinespace[-1pt] \arrayrulecolor{lightgray} \cmidrule(lr){1-16} \arrayrulecolor{black} \addlinespace[-1pt]
GraphC.Bert & \baseLabel
& 41 & 21 & {\bf 98} & 66 & 18 & 85 & 7 & 84 & 24 & 85 & 12 & 7 & 81 & 13 \\

\addlinespace[-1pt] \arrayrulecolor{lightgray} \cmidrule(lr){1-16} \arrayrulecolor{black} \addlinespace[-1pt]
CoTexT & \baseLabel
& 46 & 23 & {\bf 98} & 60 & 36 & 84 & 15 & 86 & 35 & 88 & 21 & 11 & 84 & 23 \\

\addlinespace[-1pt] \arrayrulecolor{lightgray} \cmidrule(lr){1-16} \arrayrulecolor{black} \addlinespace[-1pt]
SPT-Code & \baseLabel
& 49 & 27 & {\bf 98} & 57 & 34 & 84 & 15 & 87 & 31 & 90 & 21 & 14 & 84 & 21 \\

\midrule

\multirow{2}{17mm}{Code T5+} & EO
& 54 & 27 & {\bf 98} & 38 & 35 & 92 & 11 & 95 & 31 & 94 & 12 & 5 & 93 & 14 \\

& \baseLabel
& 39 & 19 & {\bf 98} & 60 & 21 & 83 & 10 & 88 & 26 & 89 & 16 & 9 & 83 & 15 \\

\addlinespace[-1pt] \arrayrulecolor{lightgray} \cmidrule(lr){1-16} \arrayrulecolor{black} \addlinespace[-1pt]
UniXcoder & \baseLabel
& 63 & 36 & {\bf 98} & 46 & 50 & 88 & 18 & 92 & 48 & 95 & 22 & 16 & 90 & 29 \\

\addlinespace[-1pt] \arrayrulecolor{lightgray} \cmidrule(lr){1-16} \arrayrulecolor{black} \addlinespace[-1pt]

PLBART & \baseLabel
& 35 & 19 & {\bf 98} & 56 & 20 & 83 & 7 & 78 & 20 & 84 & 11 & 9 & 79 & 12 \\

\addlinespace[-1pt] \arrayrulecolor{lightgray} \cmidrule(lr){1-16} \arrayrulecolor{black} \addlinespace[-1pt]
\multirow{2}{17mm}{Code T5} & \baseLabel
& 45 & 24 & {\bf 98} & {\bf 70} & 25 & 85 & 11 & 84 & 31 & 88 & 17 & 9 & 81 & 22 \\

& 770M
& 56 & 35 & {\bf 98} & 66 & 38 & 87 & 16 & 89 & 36 & 92 & 25 & 13 & 88 & 26 \\

\midrule

StarEnc. & \standaloneLabel
& 46 & 25 & {\bf 98} & 65 & 21 & 86 & 9 & 87 & 33 & 88 & 17 & 11 & 81 & 19 \\

\addlinespace[-1pt] \arrayrulecolor{lightgray} \cmidrule(lr){1-16} \arrayrulecolor{black} \addlinespace[-1pt]

C.RankE. & \standaloneLabel
& 80 & 58 & {\bf 98} & 38 & 70 & 94 & 34 & 98 & 66 & 98 & 33 & 19 & 94 & 47 \\

\addlinespace[-1pt] \arrayrulecolor{lightgray} \cmidrule(lr){1-16} \arrayrulecolor{black} \addlinespace[-1pt]
\multirow{2}{17mm}{C.Sage-V2} & \baseLabel
& 93 & 81 & {\bf 98} & 33 & 88 & 96 & 64 & 99 & 93 & {\bf 99} & 71 & 54 & 95 & 81 \\

& 1.3B
& 94 & 82 & {\bf 98} & 33 & 89 & {\bf 97} & 73 & 99 & 97 & {\bf 99} & 75 & 62 & 94 & 82 \\

\addlinespace[-1pt] \arrayrulecolor{lightgray} \cmidrule(lr){1-16} \arrayrulecolor{black} \addlinespace[-1pt]

\multirow{2}{17mm}{CodeXE.} & \baseLabel
& 94 & 80 & {\bf 98} & 42 & 91 & 93 & 79 & 94 & 84 & 97 & 88 & 72 & 93 & 85 \\

& 2B
& 98 & {\bf 93} & {\bf 98} & 29 & 94 & 96 & 79 & 98 & {\bf 99} & 99 & 99 & 94 & 95 & 92 \\

\addlinespace[-1pt] \arrayrulecolor{lightgray} \cmidrule(lr){1-16} \arrayrulecolor{black} \addlinespace[-1pt]

Nomic E.C. & \standaloneLabel
& 95 & 83 & {\bf 98} & 37 & 86 & 96 & 60 & 99 & 87 & 99 & 87 & 72 & 95 & 76 \\

\midrule

\multirow{2}{17mm}{Granite C.} & 3B
& 43 & 24 & {\bf 98} & 52 & 19 & 84 & 10 & 87 & 31 & 89 & 17 & 12 & 87 & 19 \\

& 8B
& 38 & 19 & {\bf 98} & 50 & 17 & 84 & 8 & 85 & 28 & 88 & 14 & 9 & 85 & 17 \\

\addlinespace[-1pt] \arrayrulecolor{lightgray} \cmidrule(lr){1-16} \arrayrulecolor{black} \addlinespace[-1pt]
C.~Llama & 7B
& 44 & 24 & {\bf 98} & 53 & 22 & 85 & 10 & 90 & 37 & 91 & 18 & 13 & 88 & 17 \\

\addlinespace[-1pt] \arrayrulecolor{lightgray} \cmidrule(lr){1-16} \arrayrulecolor{black} \addlinespace[-1pt]
Qwen3 C. & 30B
& 52 & 24 & {\bf 98} & 51 & 19 & 80 & 8 & 84 & 35 & 86 & 17 & 11 & 83 & 19 \\

\addlinespace[-1pt]\arrayrulecolor{lightgray}\cmidrule(lr){1-16}\arrayrulecolor{black}\addlinespace[-1pt]
\multirow{2}{17mm}{Qwen3 E.} & 600M
& 97 & 80 & {\bf 98} & 36 & 97 & 96 & 93 & {\bf 100} & 77 & {\bf 99} & 98 & 92 & 95 & 90 \\

& 8B
& {\bf 99} & 91 & {\bf 98} & 38 & {\bf 99} & 96 & {\bf 98} & 99 & 87 & {\bf 99} & {\bf 100} & {\bf 98} & {\bf 96} & {\bf 98} \\

 \bottomrule
\end{tabular}

\caption{Precision over the datasets BigCloneBench Type-2 (\bcbiiLabel) and Type-3 (\bcbiiiLabel), CodeNet (\codenetLabel), MultiPL-E (\multipleLabel), XCodeEval (\xcodeevalLabel). Models within each group of \Cref{tab:model-overview} are sorted by parameter count.\label{tab:bench}}

\end{table}
\renewcommand{\arraystretch}{1.0} 
\section{Model evaluation}
\label{sec:isolation}
\paragraph{Experimental Setup.}
We benchmark all models on an NVIDIA DGX Spark AI workstation~\cite{nvidia_dgx_spark_datasheet_2025} dedicated exclusively to our experiments. It features an NVIDIA Blackwell GPU (GB10 Grace Blackwell Superchip) and 128 GB of unified LPDDR5x memory. The AI software stack includes CUDA 13, PyTorch 2.12.0, and cuDNN 9.19. For precise latency, we use Python's {\tt time.perf\_counter()} and enclose each inference within {\tt torch.cuda.synchronize()} barriers, so times reflect the completion of all GPU kernels by preventing CPU-GPU asynchrony. To mitigate initialisation overhead, we perform 20 warm-up runs per model.
\paragraph{Methodology.}
For each snippet $c_i \in \mathcal{C}$ from the datasets of \Cref{sec:dataset}, we vectorise it in $v_i$ via each model from \Cref{sec:model}. To isolate embedding quality from indexing artefacts, we bypass ANN techniques in favour of exact retrieval~\cite[\S3.1]{aipowered-book} via sequential scanning. Though recent work~\cite{pitfalls_vector_sigmod26} reports that models trained on diverse objectives can be sensitive to particular distance functions, in our experiments, cosine similarity consistently matched or marginally beat Euclidean distance across all configurations; we thus report only cosine similarity in \Cref{tab:bench}. This establishes an absolute-precision baseline, letting us use the Relative Distance Error (RDE) to quantify subsequent indexing loss; we defer an end-to-end recall-then-rerank analysis to future work.
For each query $q$ (\Cref{sec:dataset}), we (i) compute the cosine similarity with each target embedding $v_i$ as $S_C(q, v_i) = \frac{q \cdot v_i}{\|q\| \|v_i\|}$; (ii) sort the results in descending order; and (iii) retrieve the top-$k$ candidates to compute precision and NDCG. Precision@$k$ (P@$k$) measures the proportion of relevant results among the top-$k$ retrieved candidates, averaged over all queries. NDCG@$k$~\cite[\S8.4]{ir_book} assesses ranking quality by weighting the position of those relevant results. 
We set $k=50$ for all datasets except MultiPL-E, where $k=20$ accommodates its smaller pool of ground-truth examples (cf.~\Cref{sec:dataset}). Experiments with smaller thresholds ($k \in \{1, 10, 20\}$) yielded consistent relative rankings, so we report only the larger $k$ to maximise candidate coverage for a downstream reranker. Our NDCG evaluation (not tabulated, for brevity) correlates strongly with precision (\Cref{tab:bench}), confirming that relevant candidates consistently occupy the top positions in recall. This bodes well for future fusion architectures, in which ensembling several models could yield high-quality candidates even at low $k$.
\paragraph{Quality of the embeddings.}
Our analysis of \Cref{tab:bench} (limited to the tabulated P@k for brevity, as NDCG behaves similarly) reveals a pronounced dichotomy that undercuts the notion of a single ``best'' model. In fact, we observe a strict trade-off: state-of-the-art retrieval quality is attainable across diverse datasets, but only at the cost of severe scalability; conversely, viable scalability on massive codebases demands substantial sacrifices in precision.
On retrieval quality alone, regardless of computational cost, a distinct top tier emerges across languages: Qwen3 Embedding, Nomic Embed, CodeXEmbed, CodeSage-V2, and, to a lesser extent, CodeRankEmbed.

For Python (first two columns of \Cref{tab:bench}), Qwen3 Embedding, CodeXEmbed, Nomic, and CodeSage dominate. On CodeNet, they achieve near-perfect P@50 scores (90--100\%), led by Qwen3 Embedding and closely followed by CodeXEmbed, Nomic, and CodeSage. On xCodeEval, the scores dip slightly, with CodeXEmbed (80-93\%) edging ahead of Qwen3 Embedding (80-91\%).

For Java (the next five columns of \Cref{tab:bench}), all models achieve 98\% on BCB Type-2; we do not tabulate Type-1, as all models achieve 100\%. Elsewhere, the top models shine on CodeNet (Qwen3 Embedding >95\%, CodeXEmbed 90--95\%, CodeSage/Nomic 85--90\%, CodeRank trailing at 70\%) and MultiPL-E (all 90--97\%), yet show a surprising weakness on BCB Type-3, where precision plummets to 29--42\%, below even the lightweight models. On xCodeEval, Qwen3 Embedding dominates (>90\%), ahead of CodeXEmbed (80\%), CodeSage/Nomic (60--75\%), and CodeRank (30--40\%).

For JavaScript, top performers achieve on MultiPL-E P@20 $\ge$ 94\%. Differences surface on the harder xCodeEval, where CodeSage and CodeXEmbed lead (97--99\%), followed by Nomic and Qwen3 Embedding (87\%), with CodeRank struggling at 66\%.

For C++, MultiPL-E results for top performers are near-perfect (97--100\%). On CodeNet, Qwen3 Embedding and CodeXEmbed dominate (99--100\%), ahead of Nomic (87\%), CodeSage (75\%), and CodeRank (33\%). On xCodeEval the gap widens: Qwen (98\%) and CodeXEmbed (94\%) clearly outperform Nomic (72\%), CodeSage (54--62\%).

For C\# (last two columns of \Cref{tab:bench}), MultiPL-E performance stays uniformly strong ($\geq 93$\%). On xCodeEval, Qwen3 Embedding and CodeXEmbed again lead (>90\%), with CodeSage (82\%) and Nomic (76\%) behind and CodeRank last (47\%).
\paragraph{Inference efficiency of the models.}
Our throughput analysis (\Cref{tab:model-overview}, rightmost column) reveals a stark efficiency gap between small and large embedding models. We measure throughput in kilobytes (KB) of raw source code processed per second (before tokenisation) on a single GPU, using a 10~MB ($\approx$9,400 functions) workload from BigCloneBench. The discrepancy is severe: lightweight encoders like StarEncoder exceed 100~KB/s with sub-10~ms latency, whereas Qwen3 Coder is roughly $47\times$ slower. This gap carries prohibitive scalability costs: a back-of-the-envelope calculation shows that indexing a terabyte-scale corpus of billions of code entities, such as Stack-Edu\footnote{\url{https://huggingface.co/datasets/HuggingFaceTB/stack-edu}}, would take several months on a single GPU with StarEncoder against years with Qwen3 Coder, a severe barrier for researchers and SMEs short of a well-equipped data centre.
\paragraph{On the quality-vs-throughput trade-off.}
Lightweight models, such as SPT-Code, CodeT5, StarEncoder, UniXcoder, GraphCodeBERT, and CoTexT, deliver acceptable baseline quality while processing code far faster. They do show dataset sensitivities: notably, they hold a clear edge on Java BCB Type-3, scoring 46--70\% (CodeT5 leading), often outperforming the heavier ``top tier''. They also perform solidly on MultiPL-E across languages, generally between 80\% and 90\%.
Their limits show on complex logic benchmarks. On xCodeEval, a recognised hard case, performance collapses to roughly 30\% or below across all languages. On CodeNet, they fare a little better but stay modest: generally below 50\% for Python and often 20--40\% for Java and C++.
UniXcoder outperforms the other fast models in specific cases, namely CodeNet Python (63\%), xCodeEval Python (36\%), CodeNet Java (50\%), and xCodeEval JS (48\%), but cannot match their surprising strength on BCB Type-3.
\begin{figure}[t]
  \centering
  \includegraphics[width=.49\linewidth]{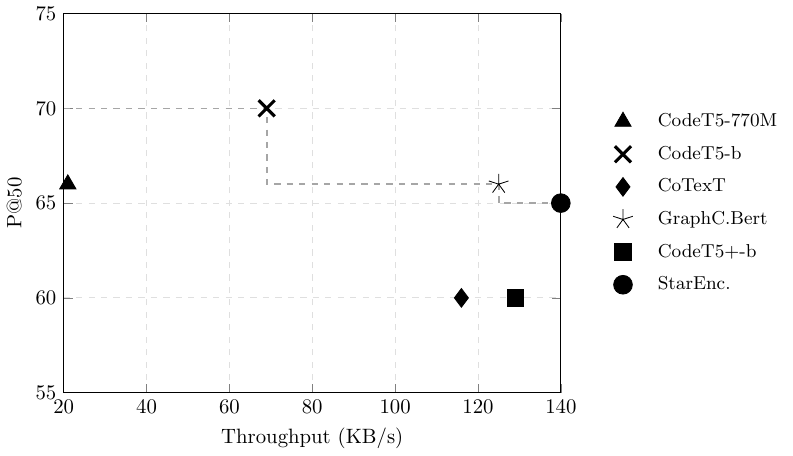}
  \caption{Throughput vs.\ P@50 over BigCloneBench (Type-3). The dashed line is the Pareto-optimal curve; the top-right corner is best.}\label{fig:scatter}
\end{figure}

As noted above, these figures expose a severe engineering bottleneck for production systems. State-of-the-art recall with models like Qwen demands an immense GPU footprint. Switching to lighter models yields roughly a 10$\times$ throughput gain, opening the door to feasible codebase indexing, but at a steep qualitative cost: P@$k$ drops are manageable on simpler tasks like MultiPL-E (only 10--15 points), yet the degradation is severe in real-world settings, with P@50 falling by 30--60 points on CodeNet and up to 80 points on xCodeEval.
Crucially, high throughput need not compromise accuracy. \Cref{fig:scatter} plots throughput against P@50 for the top models in \Cref{tab:bench}, showing that CodeT5-b, GraphCodeBERT, and StarEncoder strike the optimal precision–throughput balance on the Pareto frontier. Across most datasets and languages (omitted from \Cref{fig:scatter} for space), Qwen3-Embedder-600M consistently dominates this trade-off; its relatively low dimensionality (1024) further ensures faster distance computations and smaller index footprints.
\section{Code Rewriting}
\label{sec:rewriting}
\begin{figure}[t]
  \centering
  \includegraphics[width=.99\linewidth]{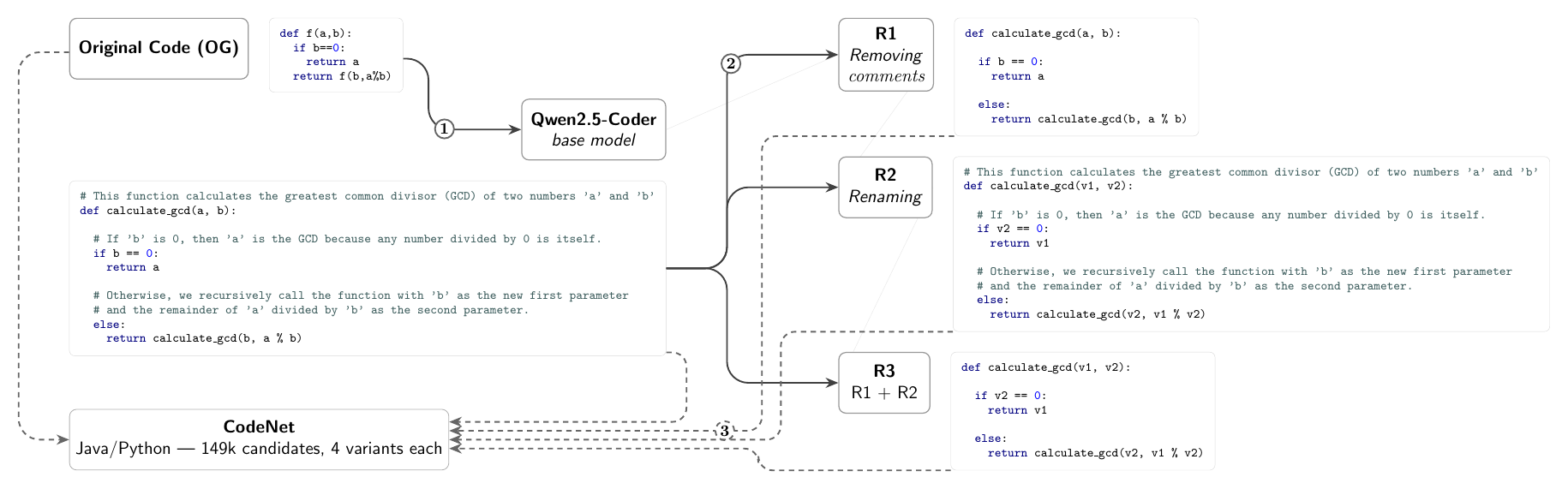}
  \caption{Rewriting scheme for variations OG, LLM, R1, R2, R3 (see \Cref{sec:rewriting}).\label{fig:rewriting-pipe}}
\end{figure}
\begin{figure}[t]
  \centering
  \includegraphics[width=.99\linewidth]{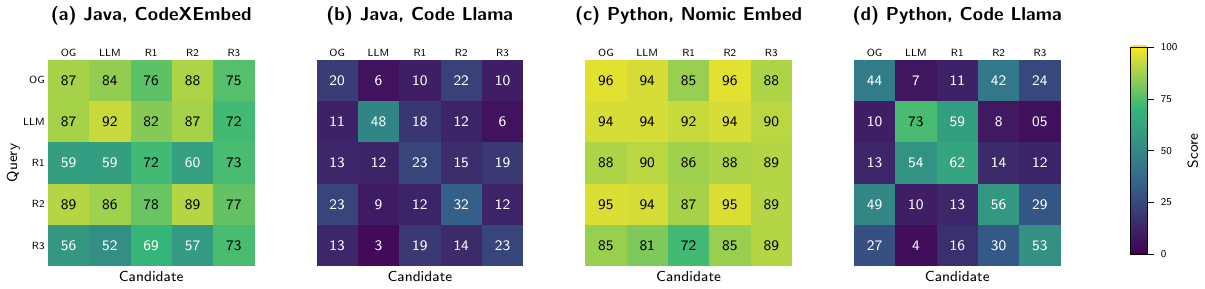}
  \caption{Query performance (P@50) for different rewriting techniques over Java and Python, and three embedding models. Rows refer to query-transformation strategies, and columns refer to codebase transformation strategies.\label{fig:heatmaps}}
\end{figure}

\cite{sacl-textual-bias,rewriting-code} show that coding style, comments, and structure influence code retrieval. MultiPL-E~\cite{multiple}, e.g., notes that models are sensitive to prompt design and comment style, though type annotations do not affect Python performance. Using the diversity of our four datasets (see Table \ref{tab:datasets}), we apply CodeNet to analyse how rewriting affects models across a wide range of performance, avoiding the ``floor'' and ``ceiling'' effects of extreme benchmarks. We evaluate four rewriting approaches for Java and Python, representative statically and dynamically typed languages, using CodeXEmbed, Code Llama, and Nomic Embed to capture diverse behavioural trends (\Cref{tab:bench}). We apply the following four code transformations with the Qwen2.5-Coder-7B-Instruct model {\em before} computing the embeddings:
\begin{description}
\item[{\sc LLM:}] Prompting Qwen2.5-Coder-7B-Instruct: ``{\it Can you please rewrite this code, adding only explanatory comments and using a cleaner programming style?}''
\item[{\sc R1:}] LLM-rewriting, then removing all comments and docstrings.
\item[{\sc R2:}] LLM-rewriting, then renaming identifiers as $v1, v2, \dots$ by appearance order.
\item[{\sc R3:}] LLM-rewriting, then both R1 and R2.
\end{description}
\Cref{fig:rewriting-pipe} illustrates the pipeline, and \Cref{fig:heatmaps} presents a selection of results for 3 models and 2 languages in $5 \times 5$ matrices, where OG denotes original code and serves as the baseline, and LLM, R1, R2, and R3 are the four transformations above. Each heatmap cell reports performance for a given model and language, with query-transformation strategies on the rows (Query) and codebase-transformation strategies on the columns (Candidate). The top-left cell (OG--OG) reports the baseline from \Cref{sec:isolation}.
\paragraph{Results.}
In Java (\Cref{fig:heatmaps} a, b), LLM-based transformation of both queries and codebase yields significant accuracy gains (cell LLM--LLM). For the top-performing CodeXEmbed, P@50 improves from 87\% (OG--OG) to 92\% (LLM--LLM); for the weaker Code Llama, it jumps from 20\% to 48\% (+28\%). This contrasts with claims that Generation-Augmented Retrieval (GAR) is ineffective~\cite{rewriting-code}, suggesting instead that style consistency eases retrieval.
Query-only rewriting in Java (leftmost column of each matrix in \Cref{fig:heatmaps} a, b) yields negligible gains. Yet when the same transformation is applied to both query and codebase (the main diagonal), the LLM approach consistently outperforms R2, particularly for weaker models, underscoring the importance of comment style in Java retrieval.
Python results (\Cref{fig:heatmaps} c, d) follow a similar pattern. For Nomic Embed, LLM-based transformations of both codebase and queries yield a modest -2\% change; for Code Llama, they yield a +29\% gain (from 44\% to 73\%).
Overall, LLM-based rewriting is effective in both languages, though the nuances differ. Comments are a consistent signal in Java, as shown by the sharp drop under R1 (-15\%, from 87\% to 72\%), yet they are less critical for top-tier models in Python. The gap between LLM and R1 reinforces findings that retrievers often rely on textual features rather than deep semantics~\cite{sacl-textual-bias}. R3 performs worst; as noted in~\cite{sacl-textual-bias}, normalising all textual features disrupts the lexical correlations (docstrings, function/variable names) that models capture during training. Even so, for Code Llama (\Cref{fig:heatmaps} b), R3 still beats the baseline.
Our key finding is that LLM-based stylistic normalisation acts as a performance equaliser: it provides substantial gains for weaker models while offering diminishing returns for stronger ones. This suggests that advanced models may implicitly perform robust normalisation, whereas simpler models benefit significantly from explicit preprocessing, a vital consideration for cost-sensitive deployments.

\section{Actionable Guidelines and Future Roadmap}
\label{sec:future}
Building on our recall-stage evaluation, future research should investigate end-to-end pipeline cost dynamics. We intend to test whether pairing lightweight embedding models with complex re-rankers optimises efficiency relative to standalone massive models. To support large-scale deployment, we will examine how variations in embedding distributions affect the efficacy of indexing structures and explore lossy-compression techniques, such as vector quantisation and knowledge distillation, to reduce storage costs for massive code corpora.
The key takeaways are threefold: (i) no recall model is universally superior; instead, a Pareto frontier is set by the task, language, clone type, and hardware budget; (ii) domain specialisation and contrastive objectives dominate raw scale, so the optimal architecture pairs a compact recall model with a powerful re-ranker; (iii) the time to index terabyte-scale corpora is the primary barrier to scalable semantic code search, making compression and hybrid retrieval essential research directions to make those objectives affordable to GPU-limited academia and SMEs.

\paragraph{To System Designers.}
The smallest viable model should be deployed in lieu of the largest available one for the initial recall stage. As two-stage retrieval requires cheap recall and accommodates more expensive re-ranking \cite{two-stage-paradigm-2023}, index dimensionality and throughput (KB/s) must serve as the primary budget constraints. Qwen3-Embedding-600M, e.g., achieves near-optimal quality (93 on Java xCodeEval and 98 on C++ CodeNet) while operating at 74~KB/s ($\sim 10\times$ faster than 2B- and 8B-parameter alternatives) with only 1024 dimensions. A 7B-parameter, 3584-dimensional embedder instead introduces penalties: embedding latency, large storage requirements, and slow vector comparisons.
Source-code style normalisation should be used to optimise costs rather than only enhance accuracy. Preprocessing code via a cheap LLM rewrite can raise a weaker encoder's performance by 28--29 points, substituting model capacity with preprocessing in cost-sensitive deployments. Yet, normalisation must preserve functional identity; aggressive transformations that strip all comments and rename variables degrade precision. Designers must also align models with the target programming language and clone type. CodeRankEmbed, e.g., performs poorly on C++ and C\# (19 and 47, respectively), as it lacks specific training for these languages (Table \ref{tab:model-overview}). Finally, if the target workload involves near-miss clone detection, a cheap, lexical-sensitive encoder often outperforms a state-of-the-art semantic embedder. Models should never be selected on a single benchmark.

\paragraph{To Model Architects.}
Our findings show that training objectives confront scale more effectively than parameter expansion. Comparing Qwen3-Embedding-600M with Qwen3-Coder-30B illustrates this: the former is 1/50th the size yet significantly outperforms the latter (97 vs 52 on Python CodeNet, 93 vs 8 on Java xCodeEval). Generative code LLMs are Pareto-dominated: they are both slow (1–9~KB/s) and inaccurate. Contrastive retrieval training is thus the critical architectural lever, whereas using decoder-only generative models in embedding mode proves counterproductive.
Architects must also account for the BCB Type-3 inversion, where lightweight encoders outperform state-of-the-art semantic embedders on gapped clones: one explanation is that robust semantic models over-normalise representations, discarding the lexical signals needed to detect Type-3 clones; this aligns with our finding that powerful models perform implicit normalisation. Models should be evaluated across the full Type-1 to Type-3 spectrum rather than focusing only on functional equivalence. Robustness to style variations distinguishes superior models: top-performing embedders remain invariant under style, comment, and identifier perturbations, whereas weaker models show performance swings of up to 30 points. We recommend incorporating rewrite-based data augmentation during training to enforce this style invariance.

\paragraph{For Future Benchmarks.} 
As for reports, computational efficiency alongside efficacy makes the quality–throughput Pareto curve a primary evaluation metric. Prior studies often omit throughput, leading to the misleading assumption that larger models are inherently superior. Benchmarks must also avoid saturating datasets. MultiPL-E scores, e.g., cluster between 94\% and 100\% across most models, with negligible discriminative power due to a severe ceiling effect, though datasets like xCodeEval separate model capabilities.
Evaluations should decouple model and index quality. Using brute-force exact nearest-neighbour search to establish a ground truth, followed by Relative Distance Error (RDE) to quantify index loss, is methodologically superior to ``black-box'' end-to-end ANN evaluations. Data contamination must be addressed rigorously as a threat to validity. As contemporary models may have ingested common benchmarks during pre-training, authors should treat training-data opacity as a formal threat and analyse cross-dataset performance fluctuations to infer exposure to datasets.
%
%
%
\bibliographystyle{splncs04}
\bibliography{sample}
\end{document}